\documentclass[iop]{emulateapj}

\def\lapp{\ifmmode\stackrel{<}{_{\sim}}\else$\stackrel{<}{_{\sim}}$\fi}
\def\gapp{\ifmmode\stackrel{>}{_{\sim}}\else$\stackrel{>}{_{\sim}}$\fi}
\usepackage{multirow}
\usepackage{color}

\begin{document}

\title{NUSTAR OBSERVATIONS OF MAGNETAR 1E~1841$-$045}

\author{
Hongjun An\altaffilmark{1}, Romain Hasco{\"e}t\altaffilmark{2}, Victoria M. Kaspi\altaffilmark{1,13},
Andrei M. Beloborodov\altaffilmark{2}, Fran{\c c}ois Dufour\altaffilmark{1}, Eric~V.~Gotthelf\altaffilmark{2},
Robert Archibald\altaffilmark{1}, Matteo Bachetti\altaffilmark{3,4}, Steven E. Boggs\altaffilmark{5},
Finn E. Christensen\altaffilmark{6}, William~W.~Craig\altaffilmark{5,7}, 
Brian W. Grefenstette\altaffilmark{8}, Charles J. Hailey\altaffilmark{2},
Fiona A. Harrison\altaffilmark{8}, Takao Kitaguchi\altaffilmark{9},
Chryssa Kouveliotou\altaffilmark{10}, Kristin K. Madsen\altaffilmark{8}, Craig B. Markwardt\altaffilmark{11},
Daniel Stern\altaffilmark{12}, Julia K. Vogel\altaffilmark{7}, and William W. Zhang\altaffilmark{11}\\}
\affil{
{\small $^1$Department of Physics, McGill University, Montreal, Quebec, H3A 2T8, Canada}\\
{\small $^2$Columbia Astrophysics Laboratory, Columbia University, New York NY 10027, USA}\\
{\small $^3$Universit{\'e} de Toulouse, UPS-OMP, IRAP, Toulouse, France}\\
{\small $^4$CNRS, Institut de Recherche en Astrophysique et Plan{\'e}tologie, 9 Av. colonel Roche, BP 44346, F-31028 Toulouse Cedex 4, France}\\
{\small $^5$Space Sciences Laboratory, University of California, Berkeley, CA 94720, USA}\\
{\small $^6$DTU Space, National Space Institute, Technical University of Denmark, Elektrovej 327, DK-2800 Lyngby, Denmark}\\
{\small $^7$Lawrence Livermore National Laboratory, Livermore, CA 94550, USA}\\
{\small $^8$Cahill Center for Astronomy and Astrophysics, California Institute of Technology, Pasadena, CA 91125, USA}\\
{\small $^9$RIKEN, 2-1 Hirosawa, Wako, Saitama, 351-0198, Japan}\\
{\small $^{10}$Space Science Office, ZP12, NASA Marshall Space Flight Center, Huntsville, AL 35812, USA}\\
{\small $^{11}$Goddard Space Flight Center, Greenbelt, MD 20771, USA}\\
{\small $^{12}$Jet Propulsion Laboratory, California Institute of Technology, Pasadena, CA 91109, USA}\\
}
\altaffiltext{13}{Lorne Trottier Chair; Canada Research Chair}

\begin{abstract}
We report new spectral and temporal observations of the magnetar 1E~1841$-$045
in the Kes~73 supernova remnant obtained with the
{\em Nuclear Spectroscopic Telescope Array (NuSTAR)}. Combined with new 
{\em Swift} and archival {\em XMM-Newton} and {\em Chandra} observations, the phase-averaged spectrum
is well characterized by a blackbody plus double power-law model, in agreement with previous,
multi-mission X-ray results. However, we are unable to reproduce the spectral results reported
using {\em Suzaku} observations. The pulsed fraction of the source is found to increase with photon
energy. The measured rms pulsed fraction is $\sim$12\%  and $\sim$17\% at $\sim$20 keV
and $\sim$50 keV, respectively. We detect a new feature in the 24--35 keV band pulse profile that is uniquely
double-peaked. This feature may be associated with a possible absorption or emission feature in the
phase-resolved spectrum. We fit the X-ray data using the recently developed electron-positron outflow model
of \citet{b13} for the hard X-ray emission from magnetars. This produces a satisfactory fit allowing a
constraint on the angle between the rotation and magnetic axes of the neutron star of $\sim$20$^\circ$ and on
the angle between the rotation axis and line-of-sight of $\sim$50$^\circ$. In this model, the soft X-ray component
is inconsistent with a single blackbody; adding a second blackbody or a power-law component fits the data.
The two-blackbody interpretation suggests a hot spot of temperature $kT\approx 0.9$ keV occupying $\sim$1\%
of the stellar surface.
\end{abstract}


\section{Introduction}

Magnetars are isolated neutron stars whose X-ray luminosities are
thought to be powered by the decay of their intense magnetic fields 
\citep{dt92, td96}.
They are observed as pulsating X-ray sources occasionally producing bright  
bursts on timescales
as short as 10 ms, as well as major enhancements in the persistent emission lasting days to months
\citep[see][for reviews]{wt06, m08, re11}.
Magnetic fields inferred from magnetar spin-down rates 
in many cases exceed $10^{14}\ \rm G$ \citep[e.g., 1E~1841$-$045, SGR~1806$-$20;][]{vg97,kds+98},
although weaker fields are suggested by recent observations of several magnetars
\citep[e.g., SGR~0418$+$5729, Swift~J1822.3$-$1606;][]{ret+10, lsk+11, rie+12, snl+12}. There are
26 magnetars discovered to date, including
candidates \citep[see][]{ok13}.\footnote{See the online magnetar catalog for a compilation
of known magnetar properties, http://www.physics.mcgill.ca/$\sim$pulsar/mag\\netar/main.html}

The X-ray spectra of magnetars often require two or more components.
The soft X-ray component (which has peak at $\sim$1 keV) is thought to be dominated by the surface 
emission from the neutron star and is likely modified by resonant scattering in the 
magnetosphere \citep{tlk02}. It may be fitted by an absorbed blackbody 
plus power law or sometimes by a two-blackbody model. The hard X-ray component 
\citep[which peaks in a $\nu F_\nu$ spectral representation above 100~keV;][]{khh+06, enm+10}
is believed to be generated in the magnetosphere. Its origin has been discussed by several authors
\citep[][]{tb05, hh05, bh07, bt07}.
Recently, \citet{b13} proposed a detailed model of hard X-ray
emission from the relativistic outflow created by $e^{\pm}$ discharge near the neutron star. 

The Galactic magnetar 1E~1841$-$045 is located at the center of the shell-type X-ray and radio supernova 
remnant Kes~73, and was first identified as an anomalous X-ray pulsar by \citet{vg97}. Its slow 
11.8-s spin period and rapid spin-down rate imply an extreme magnetic field of
$B\equiv3.2\times 10^{19}(P \dot P)^{1/2}\ \rm G=6.9 \times 10^{14}\ \rm G$, assuming 
the dipole spin-down model. Hard X-ray emission from 1E~1841$-$045/Kes~73 was detected by \citet{mcl+04},
and reported to be highly pulsed \citep[][]{khm+04, khh+06}, approaching 100\% from 15--200 keV.
Spectral studies by these authors measured a hard power-law photon index of $\sim$1.3 in the
$\sim$20--300 keV band using {\it RXTE} and {\it INTEGRAL}.  However, \citet{mks+10}
modeled the spectrum obtained with {\it Suzaku} with an absorbed blackbody plus two-power-law model
and found results only marginally consistent with the \citet{khh+06} results.

In this paper, we report on the spectral and temporal properties of 1E~1841$-$045
in the 0.5--79 keV band, measured with {\em NuSTAR}, the {\em Swift} X-ray Telescope (XRT),
{\em XMM-Newton}, and {\em Chandra}. In Section~\ref{sec:obs}, we describe 
the observations used in this paper and in Section~\ref{sec:ana}, we present the results
of our data analysis. In Section~\ref{sec:bmodel}, we apply the model of \citet{b13}
to our measurements of the hard X-ray emission from 1E~1841$-$045.
We show that our spectral fitting yields results consistent with the expectations of
the \citet{b13} model. Section~\ref{sec:disc} presents our discussion and conclusions.
These are summarized in Section~\ref{sec:concl}.

\medskip

\newcommand{\markzz}{\tablenotemark{a}}
\newcommand{\markza}{\tablenotemark{b}}
\newcommand{\markzb}{\tablenotemark{c}}
\begin{table}[t]
\vspace{-0.0in}
\begin{center}
\caption{Summary of observations used in this work
\label{ta:obs}}
\vspace{-0.05in}
\scriptsize{
\begin{tabular}{cccccccc} \hline\hline
Observatory	&Obs. ID	& Obs. Date	& Exposure	& Mode\markzz	\\ 
		&		& (MJD)		& (ks)		&	\\ \hline 
{\em Chandra}\markza	& 730		& 51754		& 10.5		& CC \\ 
{\em XMM-Newton}& 0013340101	& 52552		& 3.9		& FW/LW   \\ 
{\em XMM-Newton}& 0013340102	& 52554		& 4.4		& FW/LW   \\ 
{\em Chandra}\markzb	& 6732		& 53946		& 24.9		& TE   \\ 
{\em Swift}	& 00080220003	& 56240		& 17.9		& PC  \\ 
{\em NuSTAR}	& 30001025002	& 56240		& 48.6		& $\cdots$  \\ \hline 
\end{tabular}}
\end{center}
\vspace{-1.0mm}
$^{\rm a}${PC: Photon Counting, TE: Timed Exposure, FW: Full Window, LW: Large Window, CC: Continuous Clocking.
 MOS1,2/PN for {\em XMM-Newton}.}\\
$^{\rm b}${Used only for 1E~1841$-$045.}\\
$^{\rm c}${Used only for Kes~73 because of pile-up.}\\
\vspace{-3.0 mm}
\end{table}
\section{Observations}
\label{sec:obs}

{\em NuSTAR} is the first satellite mission that has
focusing capability above $\sim$10 keV \citep{hcc+13}.
It is composed of two focusing optics \citep{hab+10} and two CdZnTe focal plane modules \citep[][each focal plane
module has four detectors]{hcm+10},
where we use the terms ``module A'' and ``module B'' when referring to individual optics/focal plane detector sets.
The observatory operates in the 3--79 keV band with FWHM energy resolution of 400 eV at 10 keV, angular resolution
of 58$''$ (HPD, 18$''$ FWHM), and temporal resolution of 2~$\mu$s \citep[see ][for more details]{hcc+13}.

We began observing 1E~1841$-$045 with {\em NuSTAR} on 2012 November 9 at UT 22:00:02.184 with a total net exposure
of 48.6 ks.
Although {\em NuSTAR} is extremely sensitive in the hard X-ray band,
a simultaneous 18-ks {\em Swift} XRT
observation (PC mode) was conducted at UT 21:49:38.742 on 2012 November 9 to extend the spectral coverage
down to $\sim$0.5 keV where the thermal component is dominant. %
A bright point source and extended emission out to $\sim$2$'$ in radius were clearly detected
at a position consistent with that of 1E~1841$-$045/Kes~73 (see Fig.~1).

\begin{figure}
\centering
\hspace{-12.0 mm}
\begin{tabular}{c}
 \includegraphics[width=3.2 in]{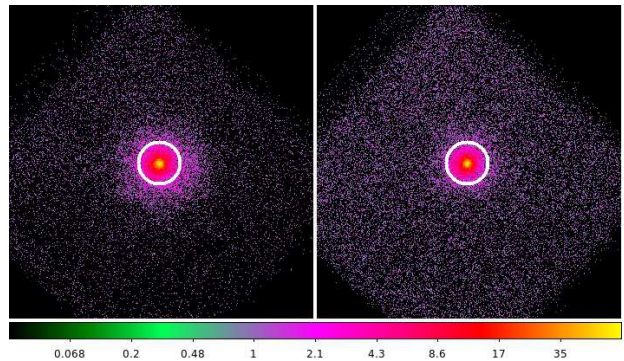}
\end{tabular}
\hspace{-15.0 mm}
\figcaption{{\em NuSTAR} images of 1E~1841$-$045 in the 3--7 keV (left) and the 7--79 keV band (right) in logarithmic scale.
1$'$ radius circles are shown in white. Energy bands were chosen such that two images have similar number of events
in the 1$'$ circle, and the scale underneath the plots shows the number of events per pixel.
Note that the diffuse Kes~73 emission (R$\sim$2$'$) is visible in the low-energy image but not
in the high-energy one.}
\vspace{2.5 mm}
\label{fig:images}
\vspace{-2.5 mm}
\end{figure}

The {\em NuSTAR} data were processed with {\ttfamily nupipeline} 1.1.1 along with CALDB version 20130509,
and the {\em Swift} data with {\ttfamily xrtpipeline} along with the HEASARC
remote CALDB\footnote{http://heasarc.nasa.gov/docs/heasarc/caldb/caldb\_remote\_acc\\ess.html}
using the standard filtering procedure \citep{cps+05} to produce cleaned event files.
We then further processed the cleaned event files for analysis as described below.
We also analyzed archival {\em Chandra} and {\em XMM-Newton} observations to have better spectral
sensitivity at low energies ($\lapp3$ keV).
The {\em Chandra} data were re-processed
using {\ttfamily chandra\_repro} of CIAO 4.4 along with CALDB 4.5.3, and the {\em XMM-Newton} data
were processed with Science Analysis System (SAS) 12.0.1. See Table~\ref{ta:obs} for a
summary of all the observations on which we report.

\medskip

\section{Data Analysis and Results}
\label{sec:ana}

\begin{figure*}
\centering
\begin{tabular}{ccc}
\hspace{-5mm}
\vspace{-5mm}
 \includegraphics[width=2.4 in]{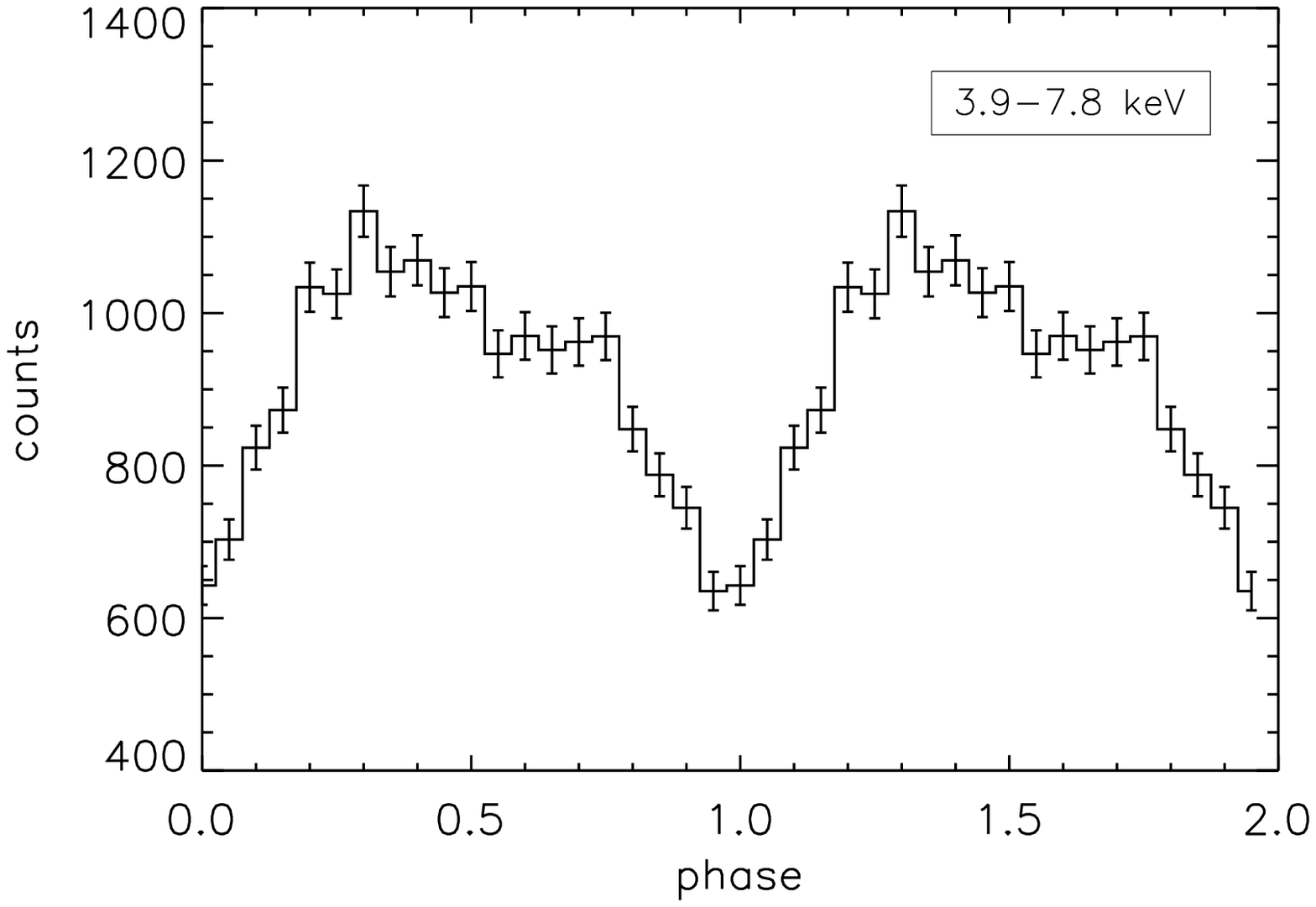} &
\hspace{-8mm}
 \includegraphics[width=2.4 in]{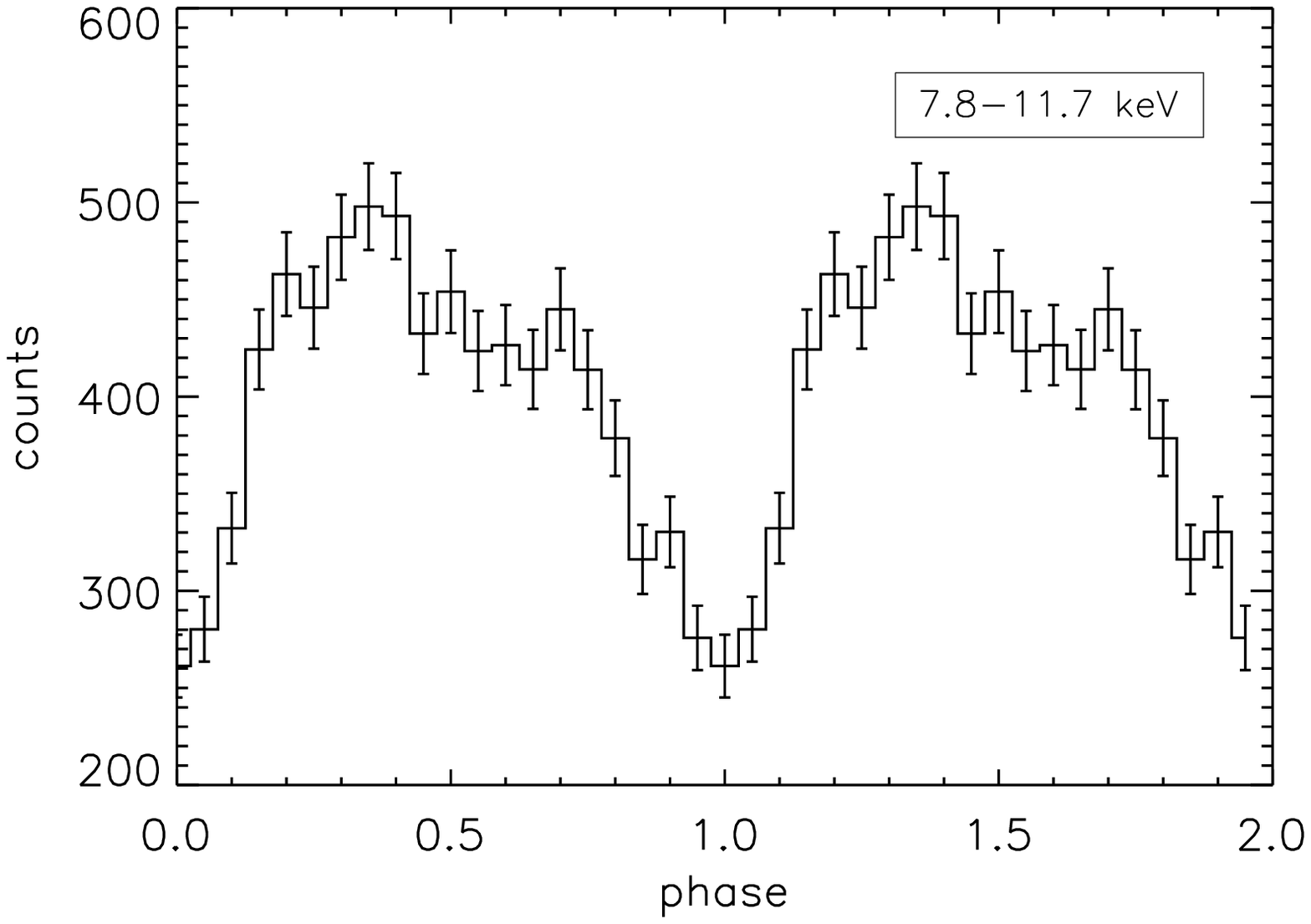} &
\hspace{-8mm}
 \includegraphics[width=2.4 in]{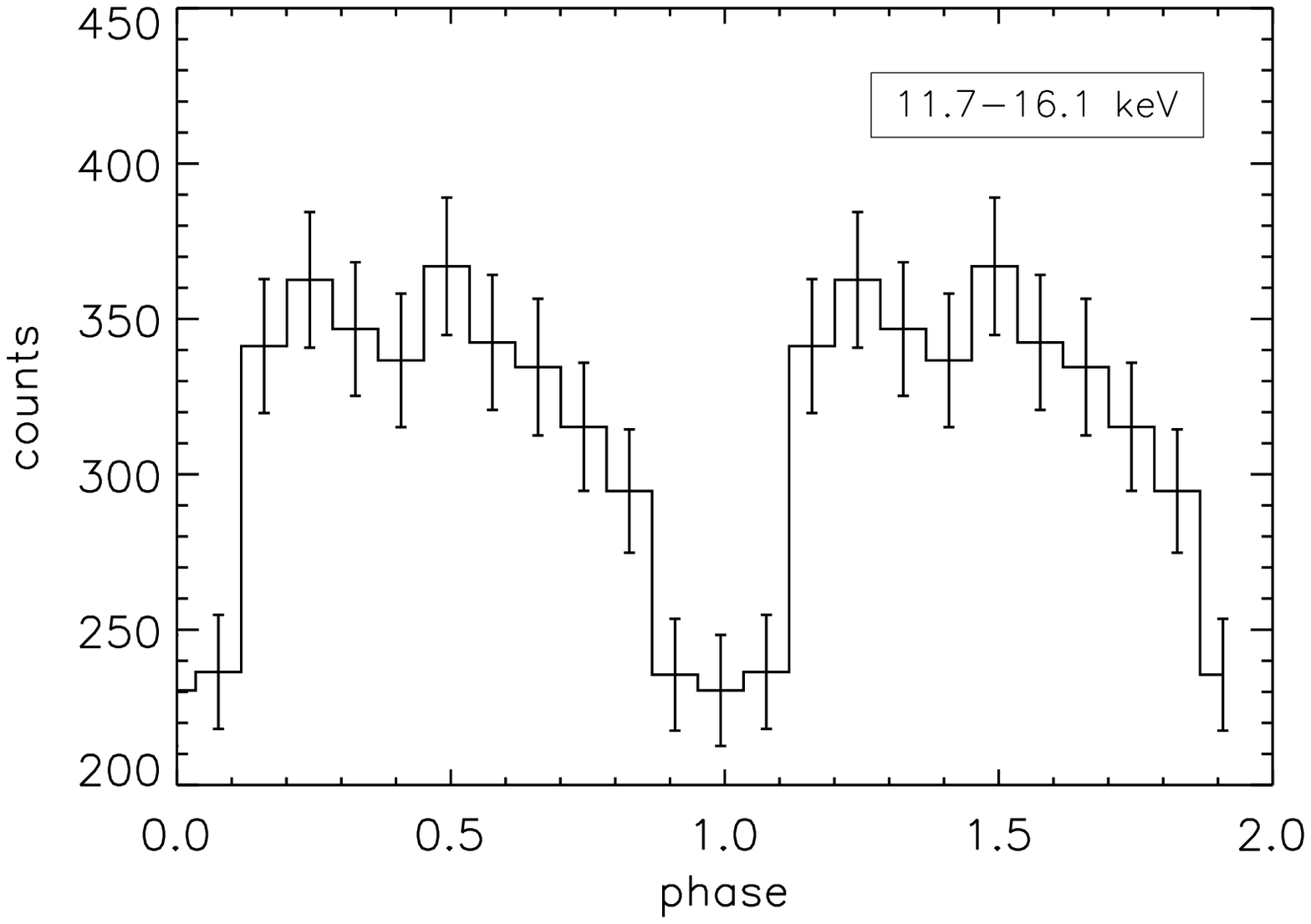} \\
\hspace{-5mm}
\vspace{-3mm}
 \includegraphics[width=2.4 in]{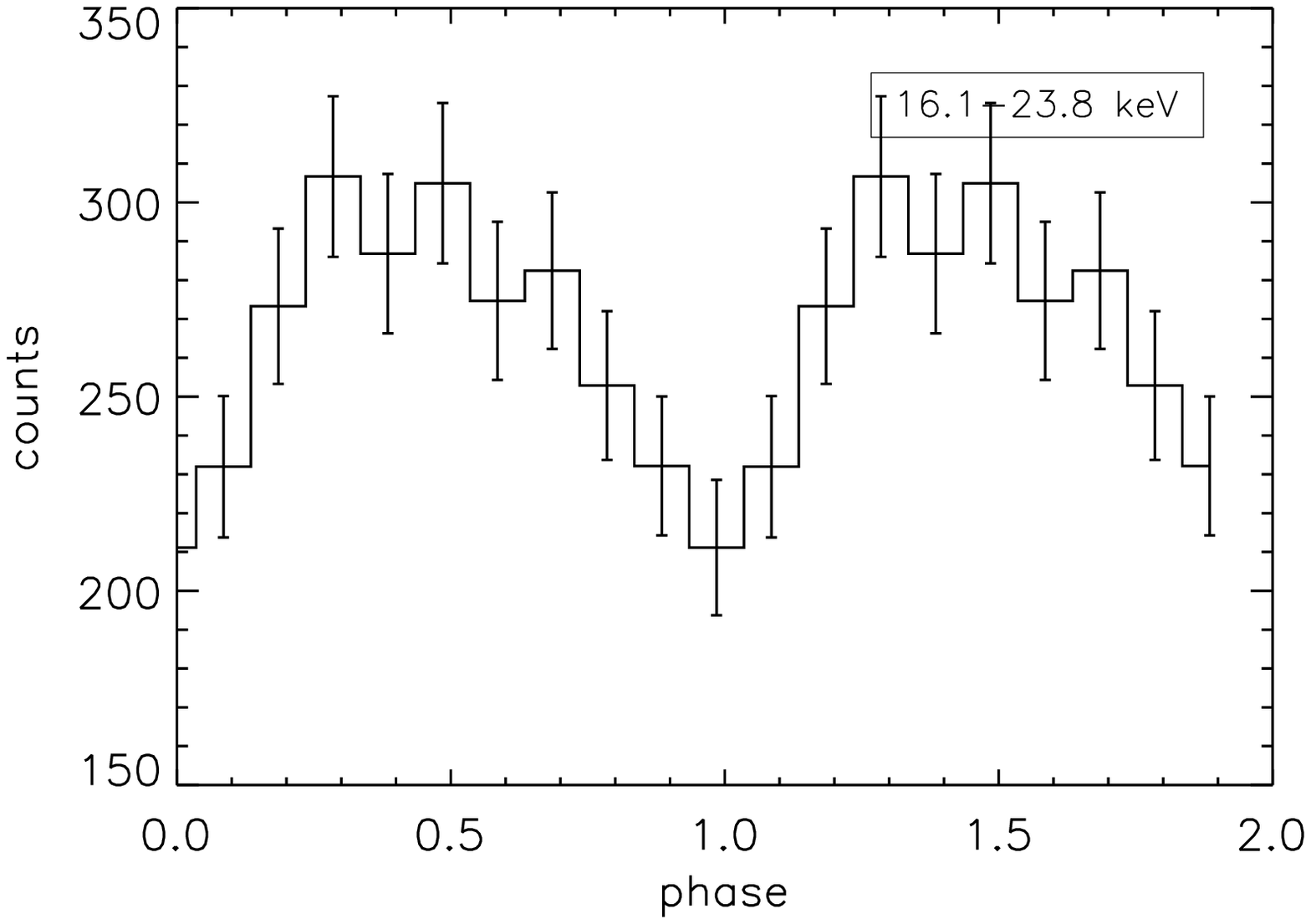} &
\hspace{-8mm}
 \includegraphics[width=2.4 in]{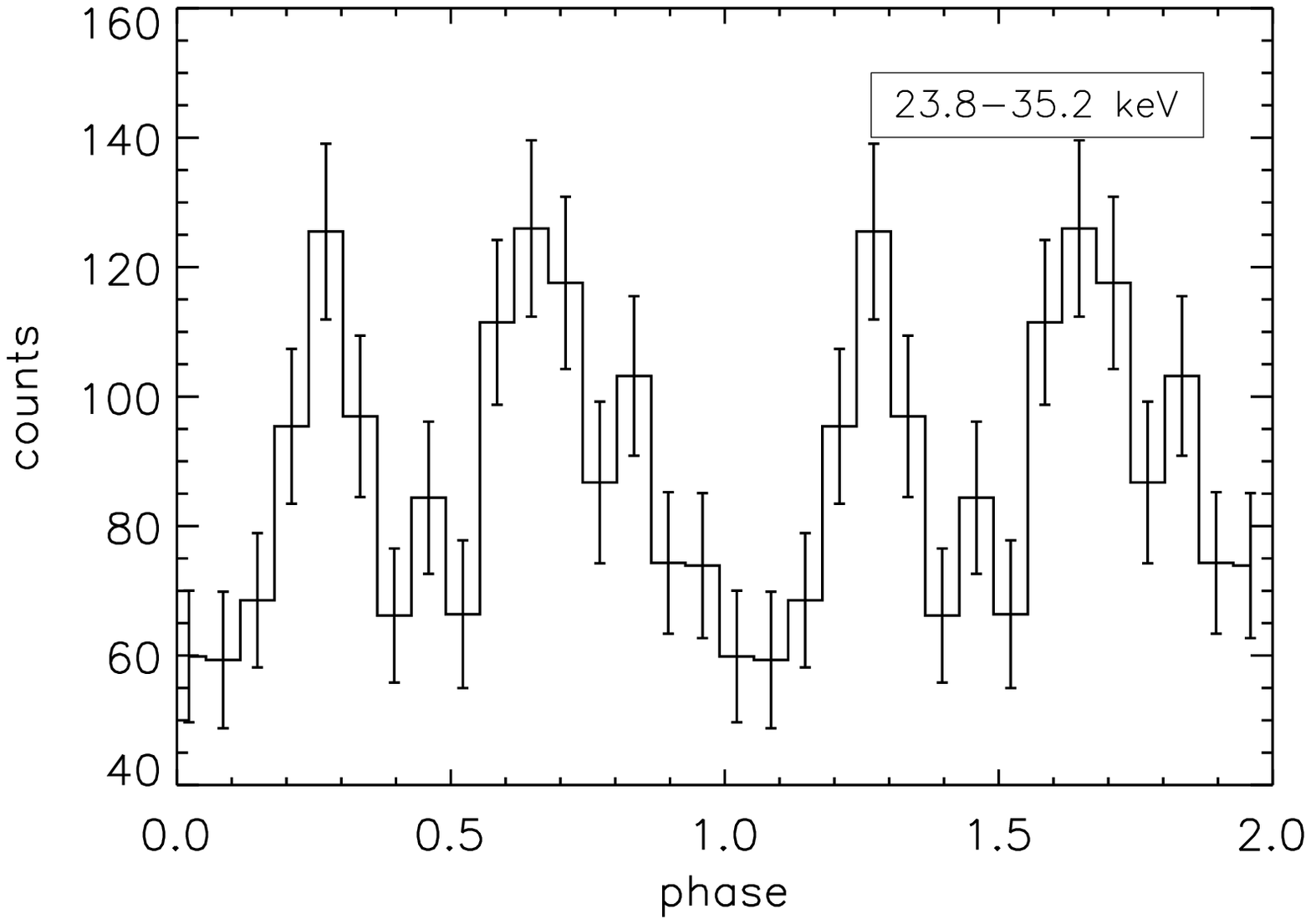} &
\hspace{-8mm}
 \includegraphics[width=2.4 in]{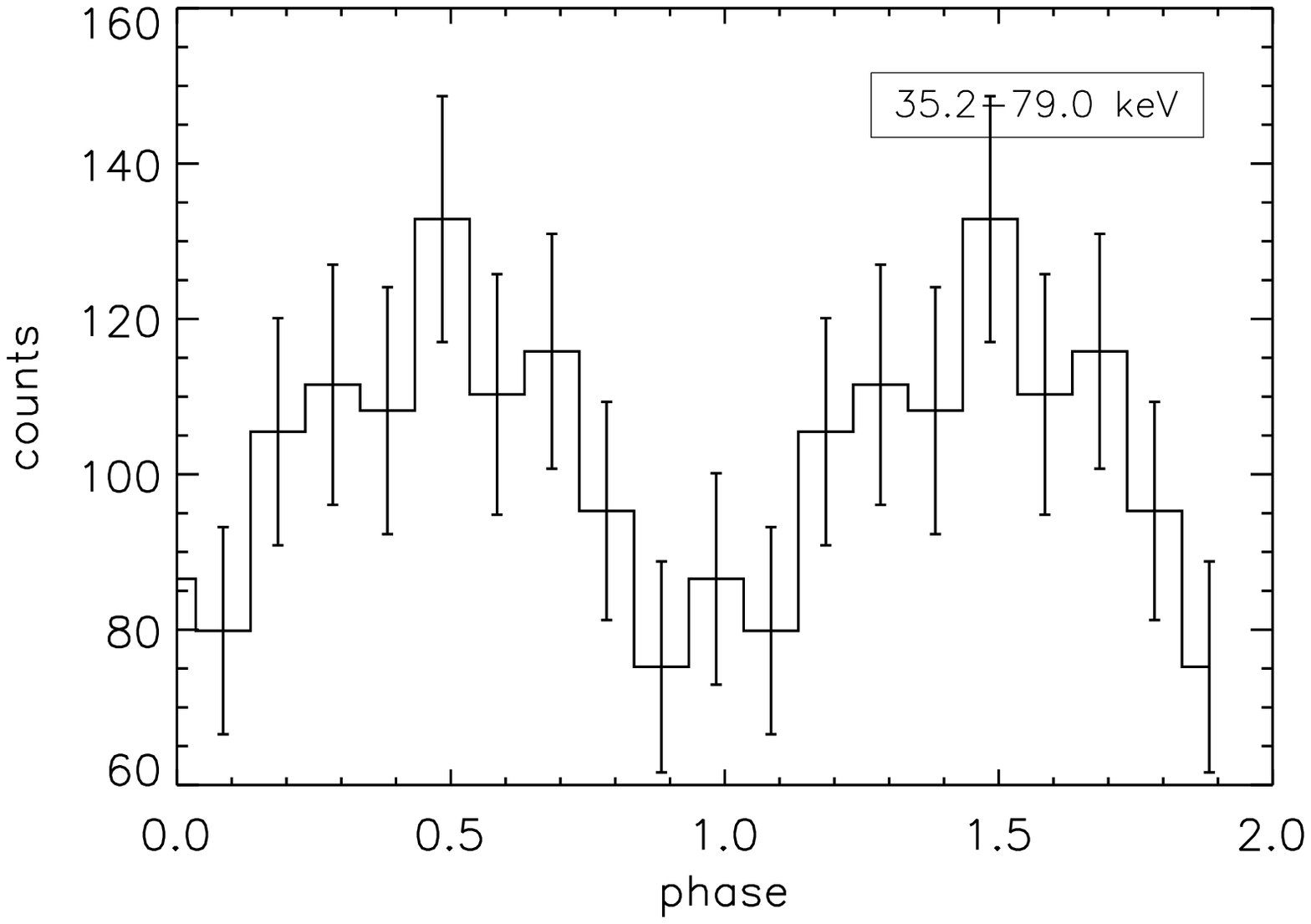}
\end{tabular}
\figcaption{Pulse profiles for 1E 1841$-$045 from {\it NuSTAR} data in various energy bands.
Note that the y-axis labels differ in each plot.
\label{fig:pulseprofile}
}
\vspace{0mm}
\end{figure*}

\begin{figure}
\vspace{-2.5 mm}
\centering
\hspace{-5.0 mm}
\vspace{2.5 mm}
\begin{tabular}{c}
 \includegraphics[width=2.5 in, angle=90]{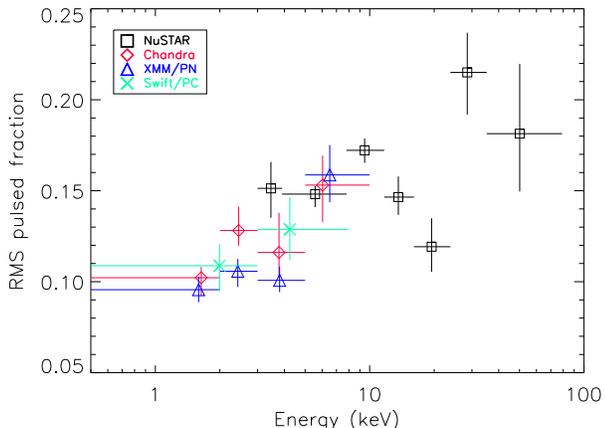} 
\end{tabular}
\hspace{-15.0 mm}
\vspace{-8.00 mm}
\figcaption{Rms pulsed fraction at several energy bands measured with four X-ray telescopes.
Note that the data point at $\sim$30 keV corresponds to the double-peaked structure in the
pulse profile.
\vspace{-0.5 mm}
\label{fig:pulsedFrac}
}
\end{figure}

\subsection{Timing Analysis}
\label{timingana}
We extracted source events in the 3--79 keV and 0.5--10 keV band within circular
regions with radii 60$''$ and 20$''$ for {\em NuSTAR} and {\em Swift}, respectively,
and applied a barycentric correction to the event lists using
the {\ttfamily barycorr} tool with the orbit files and the clock correction files using the position
reported by \citet{wpk+04}. We then used
the $H$-test \citep{dsr+89} to
search for pulsations and to measure the period.
Pulsations were detected with very high significance, and the best measured
periods ($P$) were 11.79130(2) s and 11.7914(2) s for {\em NuSTAR} and {\em Swift}, respectively.
The uncertainties were estimated using the method given by \citet{rem02}.
The periods we measured are consistent with those predicted on the basis of the ephemeris obtained
with the {\em Swift} monitoring program that will be described elsewhere
(Archibald, R. et~al. in preparation).

Since we are measuring properties of
1E~1841$-$045, the Kes~73 background must be subtracted; to do this optimally, the background region
should be within the remnant which is extended out to 120$''$ in radius from the neutron star.
Extracting backgrounds from a magnetar-free region within Kes~73 was
straightforward in the {\em Swift} data thanks to the XRT's good angular resolution; the backgrounds were extracted
from an annular region with inner radius 60$''$ and outer radius 85$''$. However, extracting backgrounds was not
easy for the {\em NuSTAR} data since the PSF is broad, and finding a source-free region within
the remnant was not possible. Therefore, we extracted the background with inner and outer radii
60$''$ and 100$''$, respectively, and
then corrected for the source contamination in the background region \citep[see][]{wg98}.
The correction factor was calculated with {\em NuSTAR}'s measured instrumental
PSF and estimated to be $\sim$10\% \citep{hcc+13}.

We also analyzed archival {\em Chandra} and {\em XMM-Newton} observations.
For the {\em XMM-Newton} and {\em Chandra} data, source events were extracted from a circle with radius 16$''$
and a rectangle with $\sim$3$'' \times 10''$ (CC mode, 3$''$ along the event distribution), respectively.
{\em XMM-Newton} backgrounds were extracted from events within an annular region with inner radius 48$''$
and outer radius 80$''$ centered at the source region, and {\em Chandra} backgrounds were extracted from
two rectangular regions with $\sim$5$'' \times 10''$ offset 5$''$ to each side from the source.
We then applied barycentric corrections to all the event lists for temporal analysis below.

We folded the source event time series at the best measured period to produce pulse profiles for multiple energy bands.
The background level was subtracted from these pulse profiles.
The background-subtracted pulse profiles obtained with {\em NuSTAR} are plotted in Figure~\ref{fig:pulseprofile}.
The energy bands were chosen to enable comparison with
those reported by \citet{khm+04}. For each energy band, the significance of pulsation was greater than 99\%.

The pulse profiles in Figure~\ref{fig:pulseprofile} qualitatively agree well with those reported by \citet{khm+04}.
However, we see a double-peaked pulse profile in {\em NuSTAR}'s 24--35 keV band.
The profile in this band has not been previously reported. To see if the apparent double peak occurred by
chance due to binning effects, we tried 250 different binnings by varying the zero phase.
For each trial binning, we fit the profile to two Gaussian functions, and measured the significance of each peak.
In all 250 cases, the significance was greater than 3$\sigma$ for both peaks.
Moreover, the two peaks do not disappear when the energy range was adjusted slightly (e.g., 25--40 keV).
Therefore, we conclude that the two peaks are genuine features in the light curve in this energy band.

In order to better constrain the transition energies of the feature, we produced pulsed profiles for smaller
energy bins (2 keV). The double-peaked structure is visible to the eye from $\sim$26 keV to $\sim$34 keV
although the structure seen in these individual profiles may not be statistically significant.

\begin{figure*}
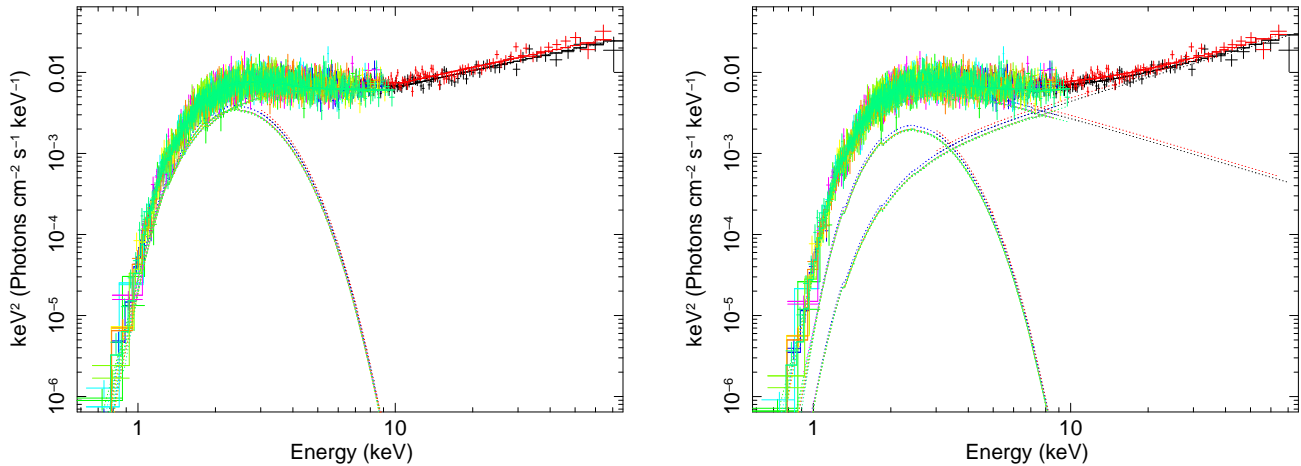

\centering
\begin{tabular}{cc}
\hspace{-7.0 mm}
 \includegraphics[width=2.4 in, angle=-90]{fig4a.eps} &
 \includegraphics[width=2.4 in, angle=-90]{fig4b.eps} \\
\end{tabular}
\figcaption{Phase-averaged spectra of {\em NuSTAR}, {\em Swift}, {\em XMM-Newton} and {\em Chandra} data.
Best-fit models and additive model components are also shown. {\it Left}: Blackbody plus broken power law,
{\it right}: Blackbody plus double power law.
\label{fig:spectra}
}
\vspace{0mm}
\end{figure*}

We calculated the rms pulsed fraction defined by
$$PF_{\rm rms}=\frac{\sqrt{2\sum_{k=1}^{4}((a_k^2+b_k^2)-(\sigma_{a_k}^2+\sigma_{b_k}^2))}}{a_0},$$
where $a_k=\frac{1}{N}\sum_{i=1}^{N}p_i \cos(2\pi ki/N)$, $\sigma_{a_k}$ is the uncertainty in $a_k$,
$b_k=\frac{1}{N}\sum_{i=1}^{N}p_i \sin(2\pi ki/N)$, $\sigma_{b_k}$ is the uncertainty in $b_k$,
$p_i$ is the number of counts in $i$th bin, $N$ is the total number of bins, and $n$ is the number of Fourier
harmonics included, in this case, $n=4$ \citep[see][for more details]{gdk+10}. We also performed similar
analyses for the {\em XMM-Newton} and {\em Chandra} data and show the measured
rms pulsed fractions in Figure~\ref{fig:pulsedFrac}. We find that the rms pulsed fraction shows a somewhat
complicated behavior with energy, and is $\sim$20\% around 50 keV but overall increases with energy.

We also searched for aperiodic variability with the {\em NuSTAR} data
in the energy band from 3--79 keV. In particular, we searched for
bursting activity in any energy band during the observations. We produced light curves with various time
resolutions (0.1--1000~s) for several energy bands (e.g., 3--79 keV, 15--79 keV, 24--35 keV).
We then searched for time bins having a significantly larger than average number
of events, accounting for the number of trials, but found none.
Therefore, we conclude that there was no bursting activity on time scales of 0.1--1000~s during
the observations.

\newcommand{\markt}{\tablenotemark{a}}
\newcommand{\markz}{\tablenotemark{b}}
\newcommand{\markv}{\tablenotemark{c}}
\newcommand{\marku}{\tablenotemark{d}}
\newcommand{\marky}{\tablenotemark{e}}
\newcommand{\markw}{\tablenotemark{f}}
\newcommand{\markx}{\tablenotemark{g}}
\newcommand{\markss}{\tablenotemark{h}}
\newcommand{\markxx}{\tablenotemark{i}}
\begin{table*}[t]
\vspace{0.0 mm}
\begin{center}
\caption{Phenomenological spectral fit results for 1E~1841$-$045
\label{ta:spec}}
\scriptsize{
\begin{tabular}{ccccccccccccc} \hline\hline
Phase & Data\markt& Energy & Model\markz & $N_{\rm H}$ 	& $kT$	&  $\Gamma_{\rm s}\markv$	& $E_{\rm break}/F_{\rm s}$\marku& $\Gamma_{\rm h}\marky$& $F_{\rm h}$\markw & $L_{\rm BB}$\markx & $\chi^2$/dof\\ 
&& (keV) & & ($10^{22}\ \rm cm^{-2}$) 	& (keV)	&		&  (keV)	&		&  &  &	 \\ \hline
0.0--1.0&S& 0.5--10 &BB+PL & 2.23(25)& 0.46(5)& 1.76(39) & 1.73(19) & $\cdots$  & $\cdots$ & 1.58(29) & 177/182 \\
0.0--1.0&X,C& 0.5--10 & BB+PL & 2.26(5)& 0.42(1)& 2.07(7) & 1.74(5) & $\cdots$  & $\cdots$ & 1.61(8) & 1866/1849 \\ \hline
0.0--1.0&N,S,X,C& 0.5--79 & BB+BP	& 2.24(4)& 0.44(1) & 2.09(4)	&  10.7(4)	& 1.33(3) & 6.84(6) & 1.91(8) & 2440/2371 \\
0.0--1.0&N,S,X,C& 0.5--79 & BB+2PL& 2.58(10)& 0.42(1) & 2.96(18)&  1.55(2) & 1.06(9)	& 5.70(9)  & 1.24(21) & 2427/2371 \\ \hline
0.15--0.5   &N,X,C& 0.5--79 & BB+BP& 2.24\markss	& 0.44(1) & 1.98(4)	&  12.4(9)	& 1.35(6)& 7.50(7) & 2.34(11) & 819/797 \\ 
0.5--0.8 &N,X,C& 0.5--79 & BB+BP& 2.24		& 0.44(1) & 1.99(5)	&  12.6(8)	& 1.18(7)& 7.78(9)  & 1.95(11) & 687/652 \\ 
0.8--1.15 &N,X,C& 0.5--79 & BB+BP& 2.24		& 0.45(1) & 2.15(6)	&  10.0(5)	& 1.27(5) & 5.79(7) & 1.69(13) & 606/633 \\  \hline
0.15--0.5   &N,X,C& 0.5--79 & BB+2PL& 2.58	& 0.42(2) & 2.99(13)	& 1.68(4) & 1.19(10)& 5.87(14)  & 1.51(26) & 816/797 \\ 
0.5--0.8   &N,X,C& 0.5--79 & BB+2PL& 2.58	& 0.45(4) & 3.04(11)	& 1.51(4) & 1.05(9)& 6.45(16) & 0.86(23) & 680/652 \\ 
0.8--1.15 &N,X,C& 0.5--79 & BB+2PL& 2.58	& 0.44(2) & 2.91(11)	& 1.37(4)& 0.91(11)&  4.89(14) & 1.33(22) & 607/633 \\  \hline
Pulsed\markxx&X,C& 0.5--10 &  PL & 2.24 & $\cdots$ & $\cdots$&  $\cdots$	& 2.40(15)& 0.43(4) & $\cdots$ & 172/299 \\
Pulsed&N,X,C& 0.5--79 &  PL & 2.24 & $\cdots$ & $\cdots$&  $\cdots$	& 1.98(7)& 1.31(6) & $\cdots$ & 429/640 \\
Pulsed&N& 5--79 &  PL & 2.24 & $\cdots$             & $\cdots$&  $\cdots$	& 1.70(12)& 1.58(15) & $\cdots$& 163/238 \\
Pulsed&N& 10--79 & PL & 2.24 & $\cdots$             & $\cdots$&  $\cdots$	& 1.36(23)& 1.72(22) & $\cdots$& 79/114 \\ 
Pulsed&N& 15--79 & PL & 2.24 & $\cdots$             & $\cdots$&  $\cdots$	& 0.99(36)& 1.76(27) & $\cdots$& 45/64 \\ \hline
\end{tabular}}
\end{center}
\vspace{-1.0 mm}
\footnotesize{{\bf Notes.} Uncertainties are at the $1\sigma$ confidence level. When combining data from different observatories, cross-normalization
factors were used. The cross-normalization factors was set to 0.9 for module A of {\em NuSTAR} (see text),
or 1 for {\em XMM-Newton} if {\em NuSTAR} data were not included. Fluxes were absorption-corrected and measured using the {\ttfamily cflux} model in
{\ttfamily XSPEC}.\\}
$^{\rm a}${ N: {\em NuSTAR}, S: {\em Swift}, X: {\em XMM-Newton}, C: {\em Chandra}.}\\
$^{\rm b}${ BB: Blackbody, PL: Power law, BP: Broken power law.}\\
$^{\rm c}${ Photon index for the soft power-law component.}\\
$^{\rm d}${ Break energy for the broken power-law (BB+BP) fit or soft power-law flux (BB+2PL)
in the 3--79 keV band if {\em NuSTAR} data were included. Otherwise, power-law flux in the 2--10 keV band in units of
$10^{-11}$ erg cm\textsuperscript{$-$2} s\textsuperscript{$-$1}.}\\
$^{\rm e}${ Photon index for the hard power-law component.}\\
$^{\rm f}${ Flux in units of $10^{-11}$ erg cm\textsuperscript{$-$2} s\textsuperscript{$-$1}.
The values are only the power-law (hard power-law) flux in the 3--79 keV band for the BP (PL, 2PL) model
when the {\em NuSTAR} data are included, otherwise power-law flux in the 2--10 keV band.}\\
$^{\rm g}${ Blackbody luminosity in units of $10^{35}\ \rm erg\ s^{-1}$ for an assumed distance of 8.5 kpc.}\\
$^{\rm h}${ $N_{\rm H}$ for the phase-resolved and pulsed spectral analysis was frozen.}\\
$^{\rm i}${ {\ttfamily lstat} in {\ttfamily XSPEC} was used for fitting pulsed spectra, and we report L-Statistic/dof
instead of $\chi^2$/dof.}\\
\vspace{-0.1in}
\end{table*}

\subsection{Phase-Averaged Spectral Analysis}
\label{spectrumana}
We extracted the source and the background events using the same regions defined in Section~\ref{timingana}.
To see if the {\em Swift} observation was piled-up, we measured the count rate within a circle of radius
20 pixels ($\sim$47$''$).\footnote{http://www.swift.ac.uk/analysis/xrt/pileup.php}
The count rate was $\sim$0.4$\ \rm cts\ s^{-1}$. Although the count rate was not large enough
to produce pile-up, we verified by removing the bright core (2--4$''$ in radius)
and found that there was no significant spectral change and thus no pile-up.
We also analyzed archival {\em Chandra} and {\em XMM-Newton} observations in order to see if there is
long-term spectral variability in the soft band and to combine them with the {\em NuSTAR} observation to
have better spectral sensitivity in the soft band.

We first fit the {\em Swift} data alone to see if there was any spectral change in the soft band (0.5--10 keV) since
the last {\em Chandra} or {\em XMM-Newton} measurements were made $\sim$12 years ago \citep[e.g.,][]{msk+03}.
We grouped the spectrum to have at least 20 counts per bin for the fit.
We first fit the data with an absorbed blackbody plus power-law model to compare with the archival
{\em XMM-Newton} and {\em Chandra} data. The spectrum is a little harder than, but consistent with
previous results \citep[][]{msk+03} as well as with our reanalysis of {\em XMM-Newton}$+${\em Chandra}
data (see Table~\ref{ta:spec}). We also tried to fit the {\em Swift} data with the same model using
the best-fit parameters obtained from modelling of the {\em XMM-Newton} and {\em Chandra} data,
and found that the {\em Swift} spectrum is consistent with the model.
Therefore, all four observations can be combined with the {\em NuSTAR} observation,
and we report fit results for the combined data.
Note that for the blackbody luminosities reported in Table~\ref{ta:spec}, we assumed a distance of 8.5 kpc based on
H {\scriptsize I} absorption measurements of \citet{tl08}.

We then tied all the model parameters between {\em NuSTAR}, {\em Swift}, {\em XMM-Newton}, and {\em Chandra}
except for the cross-normalization factors (set to 0.9 for {\em NuSTAR}; the PSF correction factor)
to fit the broadband spectrum (0.5--79 keV).
To fit the data, we grouped the spectra to have at least 100 and 20 counts per bin for
{\em NuSTAR} and the soft-band instruments ({\em Swift}, {\em XMM-Newton}, and {\em Chandra}), respectively.
We tried to fit the data  simultaneously with a blackbody plus power-law model.
In fitting, we used the 0.5--10 keV and the 3--79 keV data for the soft-band instruments and {\em NuSTAR},
respectively. The model is unacceptable ($\chi^2$/dof=2634/2373), and adding one more component improves the fit significantly.
Therefore, we fit the data to an absorbed blackbody plus broken power-law,
{\ttfamily tbabs*(bbody + bknpow)}, or an absorbed blackbody plus two power laws, {\ttfamily tbabs*(bbody+pow+pow)}
in {\ttfamily XSPEC 12.7.1}. The former is to be compared with results of \citet{khh+06} and the latter
with those of \citet{mks+10}. We note that the blackbody component was required in both models.
We show the spectra in Figure~\ref{fig:spectra} and summarize the results in Table~\ref{ta:spec}.

We studied the effects of nonuniformity in the Kes~73 supernova remnant (SNR) background because
the fit results may change depending on the background region used.
For {\em NuSTAR}, which operates in the 3--79 keV band with a relatively broad PSF, effects of the thermal
SNR and its spatial variation are likely to be very small. For other soft-band observatories, we first estimated
the background to source count rates to be $\sim$15\%, 6\% and 5\%
for the {\em Swift}, {\em XMM-Newton} and {\em Chandra} data.
Although variations on small background levels would not affect the spectral fit results
much, the background level in the {\em Swift} data was relatively high, which may be a concern.
Therefore, we used various background regions in the SNR for the spectral fits of the {\em Swift} and
{\em XMM-Newton} data. As expected from the count rate estimates,
{\em Swift} results fluctuate slightly ($\sim$20--40\% of the statistical uncertainties) depending on
the background region used, but {\em XMM-Newton} results were more stable ($\sim$6--20\% of
the statistical uncertainties). We then used the various {\em Swift} and {\em XMM-Newton} backgrounds
for the joint fit of the {\em NuSTAR}, {\em Swift}, {\em XMM-Newton} and {\em Chandra} data, and found
that the spectral variations caused by different backgrounds were typically $\lapp 10$\% of
the statistical uncertainties.

We find that our best-fit parameters for the absorbed blackbody plus two power-law
model do not agree with those of \citet{mks+10}. We checked if the spectral model of \citet{mks+10} is
consistent with the {\em NuSTAR}, {\em Swift}, {\em XMM-Newton}, and {\em Chandra} data.
We found that the \citet{mks+10} best-fit parameters do not describe our data.
The null hypothesis probability was $7\times10^{-4}$ ($\chi^2$/dof=2601/2375) with a clear
trend in the residuals at high energies ($\gapp$ 15 keV).
We then varied their best parameters for the absorbed blackbody plus two power-law model
($N_{\rm H}$=2.836--2.896$\times 10^{22}\ \rm cm^{-2}$, $kT$=0.496--0.576~keV,
$\Gamma_{\rm s}$=4.39--5.59, $\Gamma_{\rm h}$=1.42--1.82)
within the uncertainties (defined as direct sum of the statistical and
systematic uncertainties to maximize the parameter space) using the {\ttfamily steppar}
command of {\ttfamily XSPEC} to see if we could find a set of parameters that is consistent
with the data.
The minimum $\chi^2$/dof was 2530/2375, implying a null hypothesis probability of $\sim$0.01,
and some of the best-fit parameters hit the limit, making the probability lower.
We then limited the fit to the 0.5--60 keV range similar to the {\em Suzaku} data and
still found that the \citet{mks+10} best-fit parameters are inconsistent with
our data. We therefore conclude that the X-ray spectrum of 1E~1841$-$045 we measured cannot be
explained with the spectral model reported by \citet[][]{mks+10}.

\subsection{Phase-Resolved and Pulsed Spectral Analyses}
\label{spectrumana2}
We conducted a phase-resolved spectral analysis for three phase intervals, 0.15--0.5, 0.5--0.8,
and 0.8--1.15 to catch distinct features in the pulse profiles (see Fig.~\ref{fig:pulseprofile} for pulse profiles).
The temporal resolutions {\em Swift} XRT and {\em XMM-Newton} MOS are comparable to the phase intervals we
use here, and spectral mixing between different phases will occur, blurring the spectral differences among
the phase intervals. Therefore, we ignored the {\em Swift} and {\em XMM-Newton} MOS data for the
phase-resolved and pulsed spectral analysis below.

We binned the {\em NuSTAR} and the soft-band instruments' spectra to have at least 50 and 20 counts per spectral bin,
respectively, and froze the cross normalizations to those obtained with the phase-averaged spectral fit.
We fit the spectra with two models: an absorbed blackbody plus broken power-law and an absorbed blackbody plus double
power-law model. We find that the spectra vary with spin phase, and the detailed variation depends on the spectral
model used. We report the results in Table~\ref{ta:spec}.

We also fit the pulsed spectrum after subtracting the unpulsed spectrum
extracted in the phase interval 0.9--1.1. The {\em Chandra} and the {\em XMM-Newton} PN data were
phase-aligned with the {\em NuSTAR} data by correlating the light curves.
Since the number of pulsed source counts per
spectral bin was small, we used {\ttfamily lstat} because the usual $\chi^2$ method may bias the results.
We then froze the cross normalizations between instruments to the
values obtained with the phase-averaged spectral fits.

There are not many events in the pulsed spectra, and a simple power-law model
can not be ruled out. However, we see rising trends in the residuals in the soft band ($\lapp$ 2 keV)
and hard band ($\gapp$ 10 keV). Also motivated by the very hard power-law component ($\Gamma$$\sim$0.7)
in the pulsed spectrum in the high energy band ($\sim$15--200 keV) reported by \citet{khh+06},
we gradually removed the soft bands from the spectral fit to see if the spectrum becomes very hard
above $\sim$15 keV, and found that indeed it does.
We also tried to fit the data using alternative statistical methods (e.g., usual $\chi^2$ method or {\ttfamily cstat}
in {\ttfamily XSPEC}), and found that the alternative methods gave similar results except for the fit
in the 0.5--79 keV band, where $\chi^2$ results were significantly different from the others.
The results are summarized in Table~\ref{ta:spec}.

We also measured spectral pulsed fractions in the hard band (defined as the ratio of pulsed and total spectra)
in order to compare to those reported by \citet{khh+06}.
We first fit the total ($\gapp$ 11 keV) and the pulsed spectra ($\gapp$ 15 keV) to single power-law models.
The total spectrum above 11 keV is consistent with what we obtained using the absorbed blackbody plus
broken power-law model (see Table~\ref{ta:spec}).
We then calculated the flux density ratio (which we refer to as ``spectral'' pulsed fraction),
and find it to be $24\pm4$\% ($41\pm18$\%)
at 20 keV (80 keV). The uncertainties were estimated by simulating both pulsed and total spectra using
the covariance matrices obtained during the spectral fitting. Using 10,000 simulations, we calculated the
flux density ratios and the standard deviation to obtain the uncertainties.

Finally, we investigated the spectral properties of the double-peaked pulse profile in the 24--35 keV band
(see Fig.~\ref{fig:pulseprofile}).
With the double-peaked structure decidedly significant (see Section~\ref{timingana}),
we searched for evidence for this structure in the spectra.
We detect a possible excess (deficit) of counts at $\sim$30 keV in the spectrum
in the phase bin 0.525--0.725 (0.325--0.525), but not in the spectra of the other phases. 
However, the continuum model alone is statistically acceptable and the existence
of an emission- or absorption-like feature cannot be clearly demonstrated with the present data.

\subsection{Spectral fits with the $e^{\pm}$ outflow model}
\label{sec:bmodel}
Next, we tested a new model proposed by \citet{b13} to explain the hard X-ray emission from magnetars.
The model envisions an outflow of relativistic 
electron-positron pairs created by pair discharge near the neutron star.
The outflow moves along the magnetic field lines and gradually decelerates as it 
(resonantly) scatters the thermal X-rays. Its Lorentz factor decreases proportionally
to the local magnetic field $B$,
\begin{equation}
\label{eq:gam}
   \gamma\approx 100\frac{B}{B_Q},
\end{equation}
where 
$B_Q=m_e^2c^3/\hbar e= 4.44\times 10^{13}$~G. This deceleration determines 
the emitted spectrum of resonantly scattered photons. 
The outflow fills the active ``j-bundle'' (an extended bundle of electric currents)
of closed magnetospheric field lines \citep{b09}.
It radiates most of its kinetic energy in hard X-rays
before the $e^\pm$ pairs reach the top of the magnetic loop and annihilate.

In a simple geometry, the j-bundle is axisymmetric and 
emerges from the polar cap around the magnetic dipole axis of the star.
In this case, the model has the following parameters:
(1) the angular size of the polar cap, $\theta_{\rm j}$,
(2) the power of the $e^\pm$ outflow along the j-bundle, $L$, 
(3) the magnetic dipole moment of the star, $\mu$,
(4) the angle between the rotation axis and the magnetic axis, $\alpha_{\rm mag}$,
(5) the angle between the rotation axis and the observer line of sight, 
$\beta_{\rm obs}$, and (6) the reference point of the rotational phase, $\phi_0$.
See \citet{b13} for more details.

To test the model against data, we designed the following two-step 
method (Hasco\"et, R. et al., in preparation).
First, we explore
the entire parameter space by fitting the phase-averaged total (pulsed+unpulsed) spectrum
and the phase-resolved pulsed spectra.  
At this step, we only consider data above 10 keV, where the outflow dominates
the observed radiation. For 1E~1841$-$045 we used three phase bins for the
phase-resolved spectra (Section~\ref{spectrumana2}).
We found that the model successfully fits the data, with the best-fit 
$\chi^2/\mathrm{dof} = 1.13$ (for 267 dof); the obtained parameters of the 
model are given in Table~\ref{ta:bmodel}. For the best-fit model, the
spectrum ($\nu F_\nu$) peaks at $\sim$7 MeV. We also found 
a marginally acceptable ($3\sigma$ confidence) second minimum
($\chi^2/\mathrm{dof} = 1.22$ for 267 dof).
Both acceptable regions are well localized in the parameter space, and we
show both solutions in Table \ref{ta:bmodel}.

At the second step, we freeze the best-fit parameters of the outflow model 
and fit the spectrum in the 0.5--79 keV band, using the
{\em NuSTAR}, {\em Swift}, {\em Chandra} and {\em XMM-Newton} data. This allows us to analyze possible
models for the soft X-ray component. We found that the data exclude the 
single-blackbody model.
On the other hand, the data are well fitted by a blackbody plus power-law or by a two-blackbody 
model. The results are summarized in Table~\ref{ta:bmodel_fit}.
Note that the outflow model spectrum extends down
to low energies, and thus the soft-band spectral parameters are different from those
obtained using the phenomenological models (see Table~\ref{ta:spec}).
\medskip

\newcommand{\markaa}{\tablenotemark{a}}
\newcommand{\markab}{\tablenotemark{b}}
\begin{table}[t]
\vspace{0.0mm}
\begin{center}
\caption{
Best-fit parameters of the outflow model}
\vspace{-1.5mm}
\label{ta:bmodel}
\scriptsize{
\begin{tabular}{cccccc} \hline\hline
Solution & $\alpha_{\rm mag}$ & $\beta_{\rm obs}$ & $\theta_{j}$ & $L$\markaa & $\mu$\markab \\ 
	 &    (rad)           & (rad)		& (rad)		& 		&		 \\ \hline
1	 & 0.3(2) & 0.9(2)		& $<0.4$	 & 5(1)	& $>1.4$  \\
2	 & 0.7(2)	      & 1.4(1)		& $<0.4$	&  5(1)	& $>1.4$  \\ \hline
\end{tabular}}
\end{center}
\vspace{-2.5mm}
$^{\rm a}$Outflow power in units of $10^{36}(D/8.5\ \rm kpc)^2\ \rm erg\ s^{-1}$.\\
$^{\rm b}$Magnetic dipole moment in units of $10^{32}\ \rm G\ cm^{3}$.
\vspace{0.0mm}
\end{table}

\section{Discussion}
\label{sec:disc}
We have reported on X-ray observations of the magnetar 1E 1841$-$045,
most notably on its high-energy X-ray properties as observed by {\em NuSTAR}.
We find that the pulse profile in the $\sim$24--35 keV band shows a
double-peaked structure, which has not previously been reported. We also find that
the rms pulsed fraction of the source is $\sim$20\% at $\sim$50 keV.
We show that the phase-averaged total spectrum of 1E~1841$-$045 can be
modeled with an absorbed blackbody plus broken power-law or
an absorbed blackbody plus two power-law model.
Finally, we constrain the geometry of the source by fitting the phase-averaged and
the phase-resolved spectra with the electron-positron outflow model of \citet{b13}.

{\subsection{Pulse Profile}
\label{sec:timing}
The pulse profiles measured with {\em NuSTAR} broadly agree with those previously measured with {\em RXTE} \citep{khm+04}.
However, we note some differences. In the 7.8--11.7 keV band, the previously measured profile had a flat top,
from which \citet{khm+04} suggested that the dominance of the two pulses (one at phase $\sim$0.3 and the other at $\sim$0.7)
changes at $\sim$9 keV.
In the {\em NuSTAR} observation, the flattening occurred at 11.7--16.1 keV, implying the change
occurred at $\sim$11 keV, similar
to the location of the spectral break (see Table~\ref{ta:spec}). Temporally measuring the exact energy over which the
flattening occurs was difficult, and the difference between $\sim$9 keV and $\sim$11 keV may be marginal.

\newcommand{\marka}{\tablenotemark{a}}
\newcommand{\markb}{\tablenotemark{b}}
\newcommand{\markc}{\tablenotemark{c}}
\newcommand{\markd}{\tablenotemark{d}}
\begin{table*}[t]
\vspace{-3.0mm}
\begin{center}
\caption{Spectral fit results for the soft component of 1E~1841$-$045 using the outflow model
\label{ta:bmodel_fit}}
\scriptsize{
\begin{tabular}{ccccccccc} \hline\hline
 Model\marka             & $N_{\rm H}$                    & $kT_{1}$ & $kT_{2}$   &  $\Gamma$\markb     &    $F$\markc & $L_{\rm BB,1}$\markd         &  $L_{\rm BB,2}$\markd     & $\chi^2$/dof\\
                         & ($10^{22}\ \rm cm^{-2}$)     &  (keV)           &  (keV)                &                     &            &                                      &                             &  \\ \hline
 BB+PL           &  2.90(8)  & 0.55(2) & $\cdots$ & 3.79(11) & 0.55(4) & 1.08(10) & $\cdots$ & 2316/2272 \\
 BB+BB           &  2.03(4) & 0.45(1) & 0.90(4) & $\cdots$ & $\cdots$ & 2.15(7) & 0.65(9) & 2298/2272 \\
 BB                  &  1.72(2) & 0.57(1) & $\cdots$ & $\cdots$ & $\cdots$ & 2.15(4) & $\cdots$ & 2556/2274 \\ \hline
\end{tabular}}
\end{center}
\vspace{-3mm}
$^{\rm a}${ BB: Blackbody, PL: Power law.}\\
$^{\rm b}${ Photon index for the power-law component.}\\
$^{\rm c}${ Absorption-corrected flux of the power law in units of $10^{-11}$ erg cm\textsuperscript{$-$2} s\textsuperscript{$-$1}, in the 3--79 keV band.}\\
$^{\rm d}${ Blackbody luminosity in units of $10^{35}\ \rm erg\ s^{-1}$ for an assumed distance of 8.5 kpc.}\\
\vspace{-3.0mm}
\end{table*}

We found a double-peaked structure in the $\sim$24--35 keV band. It is not unusual for a magnetar's
pulse profiles to change with energy. For example, \citet{dkh08} and \citet{dkh+08} showed that the pulse
profiles of two magnetars, AXP~1RXS~J170849$-$400910 and AXP~4U~0142$+$61, change with energy.
Furthermore, the two magnetars have separate peaks in their pulse profiles that correspond to the
soft- and hard-band emission, respectively. It seems that the soft-emission peak leads the hard one
in phase at least for those two magnetars (when considering the pulse minimum as phase 0).
Although we could not clearly identify a hard peak at higher
energies for 1E~1841$-$045, the peak at phase $\sim$0.6 in the 24--35 keV band may be its counterpart;
our phase-resolved spectral analysis suggests this (see Table~\ref{ta:spec}). If this is correct,
1E~1841$-$045 behaves similarly to AXP~1RXS~J170849$-$400910 and AXP~4U~0142$+$61;
the soft-emission peak leads the hard one \citep[see also][for SGR~0501$+$4516]{gwk+10}.
It will be interesting to see if this trend is common in other magnetars.

The pulsed fraction of the source is known to increase with energy \citep{khh+06,mks+10} and
we confirm this (see Fig.~\ref{fig:pulsedFrac}). Furthermore, \citet[][]{khh+06} reported that
the pulsed fraction of 1E~1841$-$045 is $\sim$25\% at 20 keV, and $\sim$100\% above $\sim$100 keV.
Note that our measured rms pulsed fractions are shown in Figure~\ref{fig:pulsedFrac},
but cannot be directly compared to those reported by \citet{khh+06} because they reported a
spectral pulsed fraction. Therefore, we calculated the spectral pulsed fraction (Section~\ref{spectrumana2})
for the comparison.
We found that the spectral pulsed fractions are $24\pm4$\% at 20 keV, and $41\pm18$\% at 80 keV.
While they may be consistent with those of \citet{khh+06},
they may agree better with a reanalysis of the {\em RXTE} and {\em INTEGRAL} data
including more exposure (Kuiper, Hermsen, \& Beek, in preparation).

\subsection{Spectrum}
\label{sec:ttspecdisc}
We found that the spectral parameters of \citet{mks+10} are inconsistent with those we obtained using the
{\em NuSTAR}, {\em Swift}, {\em XMM-Newton}, and {\em Chandra} data.
It is possible that the discrepancy between our results and those found using {\it Suzaku} is due to
spectral variability in the source.
However, the source is known to be fairly stable, at least in the soft band \citep{zk10, lkg+11, dk13}.
\citet{mks+10} noted that the point spread function of {\em Suzaku} is broad
(HPD=2$'$), and it was difficult to subtract the Kes~73 background.
Indeed, they used an SNR model obtained with {\em Chandra} to estimate the Kes~73 background
instead of directly subtracting a measured background.
This may pose a problem in the soft band because the {\it Chandra} SNR model fit was not
very good, as previously noted by \citet{zk10}; residuals in the Kes~73 model fit would be attributed to
1E~1841$-$045 spectrum. We independently checked if the Kes~73 model
\citep[{\ttfamily vsedov} used by][]{mks+10} fit the {\em Chandra} and {\em XMM-Newton} data well,
and found that reduced $\chi^2$ values for the model were $\sim$1.3--2.3, leaving significant
residuals after the fit. Moreover, the number of SNR background events is estimated to be $\sim$70\%
larger than that of the source events for a circle of radius 110$''$ in the {\em XMM-Newton} data.
Therefore, any residuals in the Kes~73 model fits will be amplified unless the source extraction
region was small, which \citet{mks+10} could not do because of the large HPD of {\em Suzaku}.
Furthermore, difficulty in subtracting the high energy background (e.g., Galactic ridge
emission and CXB) in the {\em Suzaku}/HXD data could have made their analysis inaccurate.

\citet{mks+10} argued that the residuals are present only at the Kes~73 emission lines and did not
affect the continuum model of the point source. It is not clear if the residuals are really only at the
emission lines (for example, see their Fig.~2) and even if so, it is not clear that they do not
affect the results for such a complicated point source spectral model.

Both BB+BP and BB+2PL models are phenomenological, and we use
them mainly for comparison with previous data analyses. Although both provide a good 
fit to our data, a BB+BP model is more consistent with observations above 80~keV by \citet{khh+06}.
Note that our analysis results support the anti-correlation between 
$\Gamma_{\rm s}-\Gamma_{\rm h}$ and $B$ reported by \citet{kb10}, and
a correlation between hardness ratio ($F_{\rm h}/F_{\rm s}$,
ratio of hard to soft spectral component flux in the 1--60 keV band) and 
the characteristic age inferred from the spin-down rate by \citet{enm+10}.
 
We note that the soft-band spectrum measured with {\em Swift} in 2012 November is consistent with
those measured by {\em XMM-Newton} and {\em Chandra} 12 years ago. It has been suggested that the soft-band
spectrum of the source has been stable over 13 years between 1993 and 2007 \citep{zk10}.
This is in spite of numerous spin-up glitches and bursts \citep[][]{dkg08, lkg+11}. 
Our observations support this, in agreement with the results of \citet{dk13} which are based only
on the pulsed flux.

We found a hint of a spectral excess at $\sim$30 keV in the phase interval which
corresponds to the pulse peak (phase $\sim$0.6) of the 24--35 keV pulse profile
(see Fig.~\ref{fig:pulseprofile}). A hint of a spectral deficit was also found at the
same energy, but in a different phase interval (phase $\sim$0.45).
If we interpret this as a cyclotron line feature,
the inferred magnetic-field strength would be $\sim$3$\times 10^{12}\ \rm G$ for electron,
or $\sim$5$\times 10^{15}\ \rm G$ for proton.
The magnetic-field strength for the electron cyclotron line is similar to those in the zone where
the outflowing plasma radiates all its energy \citep{b13}.
Although the excess and deficit might be produced by line emission, longer observations are
required to demonstrate this.

With our 48-ks {\em NuSTAR} observation, the measurement of the pulsed spectrum 
in the hard X-ray band is not very precise. The obtained spectral slope in the 
15--79 keV range is $\Gamma=0.99\pm0.36$.
It is consistent with $\Gamma=0.72\pm0.15$ observed by
{\em RXTE} and {\em INTEGRAL}  in the 15--200 keV range \citep[][]{khh+06}.
\citet{mks+10} reported 
a different index of the pulsed spectrum, $\Gamma=2.45^{+0.20}_{-0.21}$.
Note, however, that they used a different energy band of
 0.7--25 keV, heavily weighted in the soft band ($<10$ keV), and thus more
representative of the soft-band spectrum.
When we limit our analysis to the 0.5--25 keV band with an $N_{\rm H}$ value of $2.87\times 10^{22}\ \rm cm^{-2}$
\citep[similar to those of][]{mks+10}, 
we find $\Gamma=2.19\pm0.09$, consistent with \citet{mks+10}.
Although we argued 
above that the {\em Suzaku} spectral results might have been biased
by Kes~73 contamination, the situation for the pulsed spectrum is different,
because the Kes~73 spectrum is subtracted in a model-independent way when subtracting the DC component.
Therefore the agreement with {\em Suzaku} results
for the pulsed spectrum is unsurprising.

When limiting the analysis to the 2--25 keV band with an $N_{\rm H}$ value of $2.54\times 10^{22}\ \rm cm^{-2}$
\citep[similar to those used by][]{khh+06}, 
the photon index became $2.00\pm0.08$, consistent with $1.98\pm0.02$, the value reported by \citet{khh+06}.

\subsection{Outflow model}
\label{ofmodel}
We found that
the phase-resolved spectrum of 1E~1841$-$045 is consistent with the 
model of \citet{b13}. In this model, the X-ray emission
comes from the active j-bundle filled with a relativistic $e^\pm$ outflow,
whose Lorentz factor decreases according to Equation~(\ref{eq:gam}).
The best-fit physical parameters are in agreement with theoretical expectations. 
Specifically, the active j-bundle is constrained to emerge from a polar cap of angular size 
$\theta_j\approx 0.4$, and  the outflow power is measured to be
$L\approx 5\times10^{36} (D/8.5 \ \mathrm{kpc})^2 \ \mathrm{erg \ s^{-1}}$.
Using Equation~(48) in \citet{b09}, one can estimate the 
voltage of $e^\pm$ discharge in the magnetosphere of 1E~1841$-$045.
It gives $\Phi\approx 10^{10}\,\psi^{-1}$~V,
where $\psi\sim$1 radian is the magnetic twist implanted in the j-bundle,
and we have used the magnetic moment of the  neutron star 
$\mu\approx 7\times 10^{32}$~G~cm$^3$, which was estimated from the 
spin-down rate \citep[][]{dkg08}.
The obtained voltage is in the expected range of $10^9-10^{10}$~V \citep[][]{bt07}.

The outflow power $L$ must be equal to the bolometric luminosity 
emitted in hard X-rays. The best-fit model shows that the spectrum peaks 
at $\sim$7 MeV, outside the {\it NuSTAR} energy range. The exact location of the peak
changes depending on the solution (Table~\ref{ta:bmodel}), but is still in the MeV band.
This is consistent with previous observations by {\em INTEGRAL} and {\em RXTE} \citep[][]{khh+06}.
Our analysis also gives constraints on the geometry 
of the magnetized rotator in 1E~1841$-$045 (see Table~\ref{ta:bmodel}), which may be 
tested and refined by future measurements of X-ray polarization (or radio polarization,
if the source one day becomes radio bright), or by incorporating
future modelling of the pulse profile.

We find that the hard X-ray emission from the $e^\pm$ outflow extends 
below 10~keV and must be included in the analysis of the soft X-ray component.
When this contribution is taken into account, we find that 
(1) a single blackbody does not provide a good fit for the soft X-ray emission,
(2) a two-temperature blackbody provides the best fit, and 
(3) a good fit is also provided by a blackbody plus power law model 
(Table~\ref{ta:bmodel_fit}).

The two-temperature blackbody model admits a simple physical interpretation.
The cold blackbody $kT_1\approx 0.45$~keV corresponds to the main thermal 
emission of the neutron star, and the hot blackbody $kT_2\approx 0.9$~keV 
comes from a hot spot.
The inferred emission areas of the cold and hot blackbodies are 
$\mathcal{A}_{1} \sim \mathcal{A}_{\rm NS} / 2$ and 
$\mathcal{A}_{2} \sim 10^{-2}\mathcal{A}_{\rm NS} $, 
where $\mathcal{A}_{\rm NS}$ is the area  of the neutron star surface 
(assuming radius $R_{\rm NS} = 10$~km).
Interestingly, $\mathcal{A}_{2}$ is comparable to the area of the 
j-bundle footprint  $\mathcal{A}_j\approx \pi\sin^2\theta_j R_{\rm NS}^2$.
The footprint is expected to be hotter than the rest of the stellar surface, as it 
can be bombarded by the particles flowing in the j-bundle toward the 
neutron star. Similar hot spots have been reported in some other magnetars
\citep[e.g.,][]{gh07, tem08}.

The soft X-ray component could also be modeled as a single blackbody
modified by resonant scattering in the magnetosphere \citep{tlk02}.
Such a modification may be expected from scattering by the 
decelerated, mildly relativistic $e^\pm$ pairs in the equatorial region of
magnetosphere \citep{b13}.
The effect is, however, currently difficult to model, because it is sensitive to the poorly 
understood velocity distribution of the highly opaque $e^\pm$ plasma near 
the magnetic equatorial plane. 

\medskip

\section{Conclusions}
\label{sec:concl}

We have analyzed
48-ks {\em NuSTAR} and simultaneous 18-ks {\em Swift} observations, and 
archival data from {\em XMM-Newton} and {\em Chandra} for the magnetar 1E 1841$-$045.
To compare with previous observations we fit the source spectra with two 
phenomenological models: an absorbed blackbody plus broken power-law 
and an absorbed blackbody plus two power laws.
The measured spectral parameters are consistent with those reported by
\citet{khh+06}, and the photon index in the hard X-ray
band is better constrained with the {\em NuSTAR}
data than before. However, the {\em NuSTAR} data are not consistent with the
spectral parameters reported by \citet{mks+10}.
Although it is possible that the source might have varied since the {\em Suzaku} observations,
it seems likely that an imperfect Kes~73 model
caused problems in the background subtraction of the {\em Suzaku} data.

Our measurements of the pulsed spectrum are less constraining
than but consistent with those in \citet{khh+06}. 
The pulsed spectrum is also consistent with {\em Suzaku} observations. We measured the
rms pulsed fraction to be $\sim$20\% at $\sim$50 keV. Although
the spectral pulsed fractions were not well constrained at high energies,
our results suggest that the pulsed fraction is likely to be significantly lower
than 100\% at 100 keV.

We find that the pulse profile in the $\sim$24--35 keV band
shows a double-peaked structure, which was not previously reported. 
The deviation of the pulse profile localized in a narrow energy range suggests
a possible absorption (or emission) feature in the phase-resolved spectrum.
Although we find some evidence for such a feature, it is not statistically significant
in the present data and requires deeper observations for possible confirmation.

The phase-resolved spectrum of 1E~1841$-$045 is consistent with the 
emission model of \citet{b13}. From the model fit, we obtain constraints 
on the angle between the rotation and magnetic axes of the neutron star. We also 
infer the size of the active j-bundle, the power of the $e^\pm$ flow, 
and the voltage of the $e^\pm$ discharge, all of which agree with theoretical expectations.
The results imply that the spectrum peaks at $\sim$7 MeV.
Using this model, we place constraints on geometrical properties of the magnetar that
in principle can be tested with future observations.

Using the physical model for the hard X-ray emission, we revisited the analysis 
of the soft X-ray component. We found that its phase-averaged spectrum can be fitted 
by two blackbodies, and the hot blackbody area is consistent with that expected for the 
footprint of the active j-bundle. However, we cannot rule out a power law plus blackbody for
the soft component; future deeper observations may help in this regard.

\medskip

This work was supported under NASA Contract No. NNG08FD60C, and  made use of data from the {\it NuSTAR} mission,
a project led by  the California Institute of Technology, managed by the Jet Propulsion  Laboratory,
and funded by the National Aeronautics and Space  Administration. We thank the {\it NuSTAR} Operations,
Software and  Calibration teams for support with the execution and analysis of  these observations.
This research has made use of the {\it NuSTAR}  Data Analysis Software (NuSTARDAS) jointly developed by
the ASI  Science Data Center (ASDC, Italy) and the California Institute of  Technology (USA). V.M.K. acknowledges support
from an NSERC Discovery Grant, the FQRNT Centre de Recherche Astrophysique du Qu\'ebec,
an R. Howard Webster Foundation Fellowship from the Canadian Institute for Advanced
Research (CIFAR), the Canada Research Chairs Program and the Lorne Trottier Chair
in Astrophysics and Cosmology. A.M.B. acknowledges the support by NASA grants NNX10AI72G and NNX13AI34G.
Part of this work was performed under the auspices of the U.S. Department
of Energy by Lawrence Livermore National Laboratory under Contract
DE-AC52-07NA27344.

\end{document}